\documentclass[nobm]{epl}

\title{Nonlinear electron transport in normally pinched-off\\ quantum wire}
\shorttitle{Nonlinear electron transport ...}
\author{Novoselov K. S.\inst{1} \and Dubrovskii Yu. V.\inst{1} \and
Sablikov V. A.\inst{2} \and Ivanov D. Yu.\inst{1}\thanks{E-mail:
\email{ivanovd@ipmt-hpm.ac.ru} } \and Vdovin E. E.\inst{1} \and
Khanin Yu. N.\inst{1} \and Tulin V. A.\inst{1} \and Esteve  D.\inst{3}
\and Beaumont S.\inst{4}}
\shortauthor{Novoselov K. S. \etal} \institute{
  \inst{1} Institute of Microelectronics Technology,
Russian Academy of Sciences - \\ Chernogolovka, Moscow District
142432 Russia\\
  \inst{2} Institute of Radio Engineering and Electronics,\\
Russian Academy of Sciences - Fryazino, Moscow District 141120
Russia\\
  \inst{3} Service de Physique de l'Etat Condense - CEA-Saclay,
91191 Gif-sur Yvette, France\\
  \inst{4} Department of Electronics and Electrical Engineering -
University of Glasgow, Glasgow G128QQ, United Kingdom }

\begin{document}

\maketitle

\begin{abstract}
Nonlinear electron transport in normally pinched-off quantum wires
was studied. The wires were fabricated from \chem{AlGaAs/GaAs}
heterostructures with high-mobility two-dimensional electron gas
by electron beam lithography and following wet etching. At certain
critical source-drain voltage the samples exhibited a step rise of
the conductance. The differential conductance of the open wires
was noticeably lower than $e^2/h$ as far as only part of the
source-drain voltage dropped between source contact and
saddle-point of the potential relief along the wire. The latter
limited the electron flow injected to the wire. At high enough
source-drain voltages the decrease of the differential conductance
due to the real space transfer of electrons from the wire in
\chem{GaAs} to the doped \chem{AlGaAs} layer was found. In this
regime the sign of differential magnetoconductance was changed
with reversing the direction of the current in the wire or the
magnetic field, whet the magnetic field lies in the
heterostructure plane and is directed perpendicular to the
current.  The dependence of the differential conductance on the
magnetic field and its direction indicated that the real space
transfer events were mainly mediated by the interface scattering.
\end{abstract}

Nonlinear electron transport in the one-dimensional (1D) ballistic
devices has attracted much interest just after the linear
conductance quantization ($G=n(2e^2/h)$, $n$ is an integer) was
revealed ~\cite{vwees,wharam}. The samples used in the experiments
were the split gate devices with negatively biased gates, which
produced a narrow constriction in the two-dimensional electron gas
(2DEG) formed at \chem{AlGaAs/GaAs} heterojunction. It was well
realized both theoretically ~\cite{glazman, fedirko} and
experimentally ~\cite{patel1,patel2,patel3} that in nonlinear
regime near half-plateau structures in the differential
conductance should be developed between the $n(2e^2/h)$ plateaus
when the external voltage $V_{SD}$ applied across the constriction
is comparable to the 1D sub-band energy spacing. More generally,
the exact value of the differential conductance of half-plateaus
is determined by the potential distribution along the constriction
and number of the 1D sub-bands involved into the electron transfer
processes between contacts ~\cite{kouwenh,martin}. It was
predicted also ~\cite{kelly} that when $V_{SD}~>~E_{F}$, with
$E_{F}$ being the Fermi energy in the source (emitter) contact,
the negative differential conductance could appear due to a
decrease of the probability for an electron to transmit through
the constriction. However this effect was not apparently observed
in the quantum constrictions for $V_{SD}$ as high as few tens of
mV ~\cite{patel1,kouwenh,brown}.

\begin{figure}
\onefigure{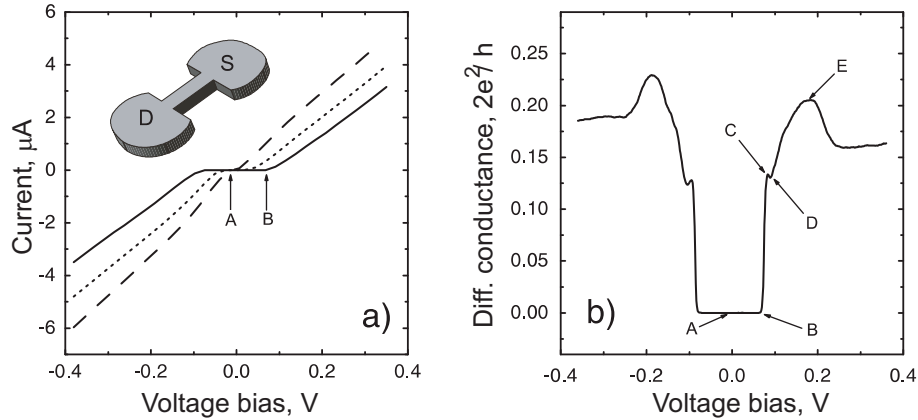}
 \caption{a) I-V characteristics of the
quantum wires with different lithographic width (dashed curve -
$0.54 \mu m$, dotted curve - $0.52 \mu m$, solid curve - $0.5
\mu m$). Insert shows the schematic draw of the samples used in the
studies. b) The differential conductance versus the voltage bias
for the sample with lithographic width $0.5 \mu m$. In the absence
of the voltage bias ($V_{SD}=0$) our sample is pinched-off (point
"A"). The step rise of the differential conductance takes place at
the critical voltage bias $V_{C}$ (point "B"). Points C, D and E
are discussed in the text.} \label{fig1}
\end{figure}

In this work the nonlinear electron transport through the
relatively long quantum wires without gates was investigated. The
wires were fabricated from \chem{AlGaAs/GaAs} heterostructure with
high mobility 2DEG by electron beam lithography and following wet
etching. The sample arrangement is shown schematically in the
insert to fig.~\ref{fig1}a. Two-dimensional reservoirs ("S" and
"D" in fig.~\ref{fig1}a) serve as contacts to the wire. Without
voltage bias $V_{SD}$ (or source-drain voltage) applied between
contacts all the wires studied were pinched off and had zero
conductance since the full electron depletion in the wire by the
built-in surface potential. At a certain critical voltage bias
$V_{C}$ the samples exhibited a step-like rise in the conductance.
The similar behaviour was observed in Ref.~\cite{kouwenh} on the
split-gate devices initially pinched-off by the gate voltage.
Under applied voltage $V_{SD}$ higher than $V_{C}$, the current
was determined by the electron flow injected to the wire through
one-dimensional sub-bands at the saddle-point located close to the
source contact. In contrast to the previous studies, we extended
source-drain voltage range up to the value as large as $0.4~V$
without sample destroy and found that at high enough voltages the
real space transfer of electrons occurs from the wire in the
\chem{GaAs} well to the doped \chem{AlGaAs} layer. This process
affects significantly the electron transport through the wire. In
this regime the sign of differential magnetoconductance was
changed with reversing the direction of the in-2DEG-plane magnetic
field transverse to the current or the current flow direction in
the wire.

The quantum wires of presented studies are narrow constrictions
(mesa structures) with the length of approximately $0.8 \mu m$ and
the width of approximately $0.4 \mu m$ fabricated by E-beam
lithography and following wet etching of the \chem{GaAs/GaAlAs}
heterostructure with high-mobility 2DEG. The starting
\chem{GaAs/GaAlAs} heterostructure has the 2DEG at 100~nm beneath
the surface (electron mobility is
$1.4\cdot10^{6}~cm^{2}V^{-1}s^{-1}$ at $T=4.2~K$ which
corresponds to $9~\mu m$ transport electron mean free path, and the
electron density - $3.1 \cdot 10^{11}~cm^{-2}$ without
illumination). The 2DEG is separated from the 50~nm thick Si-doped
($8 \cdot 10^{17}~cm^{-3}$) \chem{AlGaAs} layer by the undoped
25~nm thick spacer \chem{AlGaAs} layer.

The real width of the constrictions differed from the lithographic
width since the wet etching was used. For example, the sample with
lithographic width of $0.54 \mu m$ had the real width of about $0.4
\mu m$. Below we refer to the lithographic width when comparing
different samples.

Figure~\ref{fig1}a shows the current-voltage (I-V) characteristics
for the samples with nominal lithographic widths of $0.50 \mu m$,
$0.52 \mu m$ and $0.54 \mu m$ at $4.2~K$. Increasing bias voltage
induced a stepwise increase of the differential conductance $G =
\partial I/\partial V$ from zero to an approximately constant
value $G_{0}$ (various for all samples) at a certain critical
voltage $V_{C}$. We determined $V_{C}$ as the voltage when the
current through the sample became equal to 10~nA. The critical
voltage $V_{C}$ increases strongly with decreasing of the
lithographic width of the wire.

The differential conductance versus voltage bias for the sample
with lithographic width $0.5 \mu m$ is shown in fig.~\ref{fig1}b.
The data are similar for all the samples and were acquired by the
standard lock-in technique. One can see the local maximum (marked
by "C" in fig.~\ref{fig1}b) close to the critical voltage $V_{C}$.
At higher voltages the differential conductance reaches the
maximum (point "E" on the fig.~\ref{fig1}b) with
$G\approx0.2(2e^2/h)$, and then falls down to the approximately
constant value $G\approx0.15(2e^2/h)$.
\begin{figure}
\onefigure{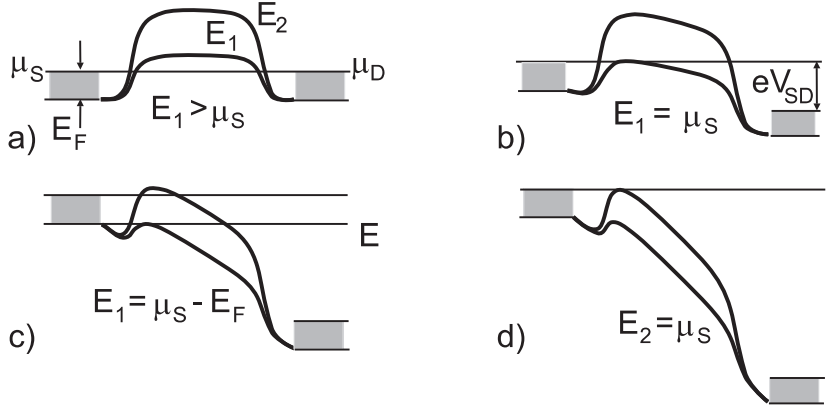}
  \caption{ Schematic view of the 1D energy diagram along the
quantum wire with contacts. The relative position of the
conduction band bottom in the contacts and the bottom of the first
and the second sub-bands in the wire are shown for different
voltage bias. $E_{1}$ and $E_{2}$ denote the maxima of the
potential barrier for the 1st and 2nd sub-bands respectively.
(a) - no source-drain voltage is applied (this corresponds to
point A in fig.~\ref{fig1}b); (b) - $V_{SD}=V_{C}$, the bottom
of the first sub-band coincides with the Fermi energy in the
emitter (point B in fig.~\ref{fig1}b); (c) - the bottom of the
first sub-band coincides with the bottom of the conduction band in
the emitter (point C in fig.~\ref{fig1}b); (d)- the bottom of
the second sub-band becomes lower than the Fermi energy in the
emitter (point D in fig.~\ref{fig1}b). } \label{fig2}
\end{figure}

The main features of the I-V characteristics can be understood
using the simple 1D model of the conduction band, shown in
fig.~\ref{fig2}. Without the bias voltage the maximum of the
ground sub-band bottom lies above the Fermi level in the 2D
contacts (fig.~\ref{fig2}a). The inter sub-band energy can be
estimated taking into account that the constriction is strongly
depleted with electrons and therefore the electric potential is
produced by the positive background charge of impurities in the
doped \chem{AlGaAs} layer. Thus the second derivative of the
potential, which determines the inter sub-band energy, is
estimated as $\phi''\sim 4\pi N_D/\epsilon_L$, where $N_D$ is the
donor concentration in the \chem{AlGaAs} layer, $\epsilon_L$ is
the lattice dielectric constant. In such a way the inter sub-band
energy is estimated as about 10 meV.

With further increase of the source-drain voltage the barrier for
the first 1D sub-band is lowered. When the applied voltage is
equal to a critical value $V_{C}$, the highest point (saddle
point) of this barrier coincides with the Fermi energy in the
source. At this moment a step-like rise of the differential
conductance must be observed (fig.~\ref{fig2}b) at zero
temperature, if the tunnelling processes are not taken into
account.

Under the voltage above $V_{C}$, the saddle point is located close
to the source and electrons are injected through this point into
the wire where the electric field is as high as about 1~kV/cm. The
electron transit time through the wire is about $10^{-12}~s$ which
is slightly more than optical phonons emission time. This argues
that electrons transit the wire near ballistically.

Under this condition the differential conductance of the structure
is determined by the sum of the partial conductances due to the 1D
sub-bands. The partial differential conductance through the open
sub-band channel $G_{n}=L_{n}(2e^{2}/h)$ is determined by the
leverage factor $L_{n}$ which is defined as
$L_{n}=V_{n}/V_{SD}$,where $V_{n}$ is the part of the source-drain
voltage which drops between the source and the top of the barrier
(a saddle point of the potential relief) of the n-th sub-band
\cite{kouwenh,martin}. In general the factor $L_{n}$ depends on
the external voltage bias: $L_{n}=f(V_{SD})$. When the applied
voltage is increased the saddle point position goes to the source.
Hence the differential conductance should decrease with $V_{SD}$.
As the voltage bias is further increased the top of the barrier
for the first 1D sub-band goes to the bottom of the conductance
band in the source 2DEG (fig.~\ref{fig2}c), and the electron flow
through the first sub-band reaches its saturation. Correspondingly
the differential conductance due to this partial current goes to
zero. The same happens with the electron flows through the higher
sub-bands. As an example, fig.~\ref{fig2}d shows the situation
where the second sub-band barrier top intersects the Fermi level
$E_{F}$.

Thus, the partial differential conductance $G_{n}$ for each
sub-band behaves with the applied voltage $V_{SD}$ as follows. It
is very small when $V_{SD}~<~(V_{C})_{n}$, at
$V_{SD}~=~(V_{C})_{n}$ the conductance increases sharply to the
value $G_{n}=L_{n}(2e^{2}/h)$ and then decreases up to zero.
$(V_{C})_n$ here is the critical voltage $V_{C}$ for the n-th
subband. The total differential conductance is a much more
complicated function of $V_{SD}$. We believe that the local
maximum observed slightly above $V_{C}$ in fig.~\ref{fig1}b is
just the first local maximum, which should appear as described
above. The local maximums of the higher order were not resolved on
our samples.
\begin{figure}
\onefigure{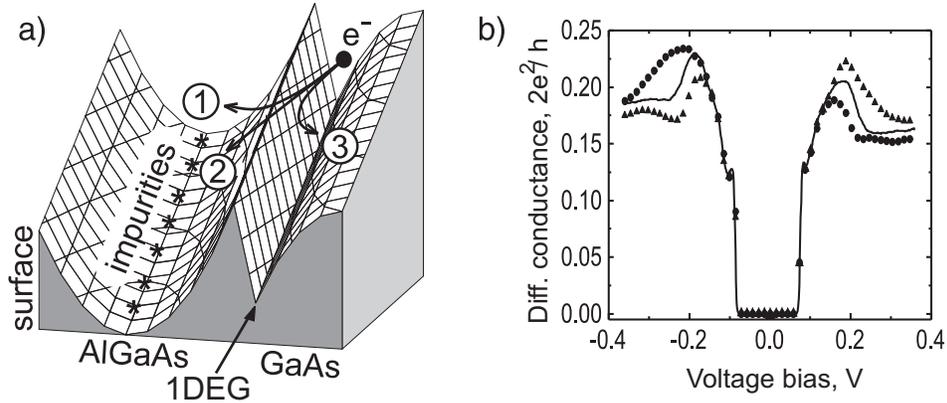}
  \caption{ a) Schematic profile of the conduction band bottom
in the heterostructure under applying a high voltage bias
($>0.2~V$). Arrow 2 shows schematically the ballistic trajectory
of injected electrons in the wire without magnetic field. Arrows 1
and 3 show the same but with the magnetic field present. The
injected electrons are forced against the interface, when the
magnetic field directed down, and deflected in the opposite
direction, when the magnetic field sign is reversed. b) The
differential conductance versus the voltage bias for the sample
with lithographic width $0.5 \mu m$ in the absence (solid curve)
and in the presence of the magnetic field. The magnetic field is
directed perpendicular to the current, in plane of the 2DEG. For
dashed and dotted curves the magnetic field is directed in the
opposite directions. } \label{fig3}
\end{figure}

Figure~\ref{fig3}a illustrates by arrow 2 how the electrons
injected through the saddle point travel along the constriction.
It is obvious that if $V_{SD}$ is high enough, the injected
electrons can acquire the energy from the driving electric field
which exceeds the height of the barrier separated the bound states
in the \chem{GaAs} well and states in the \chem{AlGaAs}. For
\chem{Al_{0.24}Ga_{0.76}As} this barrier height is about $0.2~V$.
In this case the real space transfer of electrons from the
\chem{GaAs} quantum channel into the \chem{AlGaAs} layer is
possible.

In the highly doped \chem{AlGaAs} layer, the electron scattering
is much stronger than in the \chem{GaAs} quantum well due to
presence of charged impurities and imperfections. Hence the
electron transfer to the \chem{AlGaAs} layer results in the
scattering of the injected electrons and in the increase of the
source-drain voltage fraction dropped between the saddle point and
the drain. As the result, the $L_n$ factor diminishes. This
explains the differential conductance decrease with the $V_{SD}$
voltage after the broad maximum shown in fig.~\ref{fig1}b. At
higher voltages the differential conductance only slightly depends
on $V_{SD}$.

The magnetic field effect on the electron transport in nonlinear
regime allowed us to clarify the mechanism of the real space
transfer process. It depends on the direction of magnetic field
when magnetic field vector {\bf B} lies in the plain of the 2DEG
and is directed normally to the current. This is illustrated in
fig.~\ref{fig3}b where the dependence of the differential
conductance versus the voltage bias is shown for three cases:
({\em i}) the magnetic field is absent (solid curve); ({\em ii})
the vector {\bf B}, the current vector {\bf j} and the outer
normal vector {\bf n} to the sample surface form the left-handed
system (triangles); and ({\em iii}) the vector {\bf B} is opposed
to that in the previous case (circles). The change of the
magnetoresistance sign with the change of the magnetic field or
current direction was observed when the source-drain voltage is so
high that the real space transfer of electrons from the quantum
wire to the \chem{AlGaAs} layer becomes feasible.

The similar odd magnetoresistance effect was observed previously
by Sakaki~\cite{sakaki} in the case of the diffusive electron
transport along an asymmetrical quantum well under linear
transport conditions. It was explained by the change of the
effective scattering rate by charged impurities located near the
AlGaAs/GaAs interface with the magnetic field shifted electrons
closer to or farther from the interface.

At low voltages ($V_{SD}<0.13~V$) only positive magnetoresistance
was found.

The observed decreasing of the differential conductance at high
applied voltage and its behaviour with magnetic field can be easy
understood if we suggest that the electron transitions from the
quantum wire to the \chem{AlGaAs} layer are caused by the
GaAs/AlGaAs interface scattering. The transition probability is
proportional to the matrix element $|<\chi_n|U|\psi>|^2$ where $U$
is a scattering potential localized near the interface, $\psi$ is
a wave function extended over the \chem{AlGaAs} layer, $\chi_n$ is
the wave function of electrons confined in the wire. The wave
function $\chi_n(y,z)$, $y$ and $z$ being the transverse
coordinates, is determined by the potential confining electrons in
the lateral direction ($y$) and by the triangular well produced by
the built-in electric field $F_s$ normal to the interface ($z$
direction). The magnetic field affects the wave function in the
normal direction. It changes the confining field by adding the
Lorenz force: $F = F_s + \hbar B k(x)/(mc)$, where $k(x)$ is the
wave vector describing the electron motion along the wire.  If the
Lorenz force is directed toward the interface, the subband bottoms
in the wire are shifted up and the wave function is localized
closer to the interface. Both these factors enhance the electron
transitions from the wire to the \chem{AlGaAs} layer. As a result
the magnetic field decreases the conductance.  The magnetic field
of opposite direction suppresses the real space transfer of
electrons giving rise to the conductance increase. The value of
the magnetoresistive effect is determined by the ratio of the
Lorenz force to $eF_s$. The wave vector $k$ can be estimated as
$k\sim (2meV_{\rm sd})^{1/2}/\hbar$. Under the experimental
conditions $V_{\rm sd}\sim 0.2~V$ and $B\sim 8~T$, this estimation
shows that the equivalent confining electric field is about
$2\cdot10^5~V/cm$, which is comparable with the built-in electric
field. Hence the magnetic field essentially affects the transport
of ballistic electrons in the wire and gives rise to the strong
magnetoresistive effect.

In conclusion, we have investigated the nonlinear electron
transport through relatively long quantum wire which is normally
pinched-off by the build-in surface potential. The current
transport is observed when the driving (drain-source) voltage
exceeds a critical value $V_{C}$ above which the quantum wire
becomes open. At high enough applied voltage, comparable with the
conduction band offset between \chem{GaAs} and \chem{AlGaAs}, the
real space transfer of electrons from the quantum wire into the
highly doped \chem{AlGaAs} layer becomes important. The magnetic
field, that lies in the heterostructure plane and is directed
transverse to the wire, results in a high magnetoresistive effect.
It is caused by the magnetic field influence on the real space
transfer processes. The dependence of the magnetoresistance sign
on the direction of magnetic field or current in the wire
indicates that electrons real space transfer events are mainly due
to the interface scattering.

This work was supported by the INTAS (grant 96-0721), and partly
by RFBR (98-02-17642 and 99-02-18192), and national program
"Physics of solid state nanostructures" (97-1057).

\end{document}